# Compact vortex structures' dynamics in HTS bulks with levitation techniques

Alexander A. Kordyuk*, Volodymyr V. Nemoshkalenko

Institute of Metal Physics, 36 Vernadsky St., Kyiv 03142, Ukraine

*The discovery of high temperature superconductors (HTS) has led to understanding that, in order to explain and utilize the phenomenon, completely new physical approaches should be introduced at all scales: microscopic, mesoscopic, macroscopic. Leaving first two scales beyond the scope of the present paper we focus in the upper limit of the last one, the study of the magnetic flux dynamics in HTS bulks — the dynamics of the 'compact vortex structures'. A new direction in the experimental superconducting physics, the investigation of HTS bulks with levitation techniques, which has been elaborated during last years to effectively explore the subject, as well as the new fundamental and applied results obtained therefrom are overviewed here.*

*Key words:  levitation, high-temperature superconductors, AC egergy loss, vortex dynamics, pinning, critical current density, granular HTS, mett-textured HTS*

---

* Email: kord@imp.kiev.ua, homepage: http://www.imp.kiev.ua/~kord

CONTENTS

# Introduction

The discovery of high temperature superconductors (HTS) in 1986 has revealed a number of holes in understanding of the phenomenon of superconductivity. And it is not clear up to now can the high temperature superconductivity be described in terms of usual phonon coupling at microscopic level or unconventional physics should be involved (see [1–3] and references therein), but introduction of HTS into a focus of the intensive studies has resulted in the opening (or reopening) of many other aspects of superconducting physics. The 'vortex matter' is the most known example [4, 5]. This conception spreads to the more general class of physical problems than the macroscopic magnetic properties of superconductors and, while the understanding of the vortex nature in HTS is far from to be complete [6], these materials are considered as a 'gift of nature' [7, 8] to study the 'vortex matter' in. At that, in real materials and practicable experimental geometries, there is an additional point which highly complicates the problem — non-straight vortex configurations. In case of a uniform magnetic field but non-trivial sample shapes this leads to an appearance of the so-called 'geometric barrier' [9, 10]. In case of non-uniform magnetic field, the 'compact vortex structures' bring essential peculiarities in vortex dynamics [11]. The interaction of a non-uniform magnetic field with bulk superconductors is the main subject of this review.

After the discovery of HTS, an opportunity to probe the superconducting state with liquid nitrogen induced a number of investigations of different systems with superconducting levitation which have been found to be very suitable to study the compact vortex structures dynamics in HTS and have formed a new branch in superconducting experimental physics [12, 13]. Moreover, after the development of the melt-textured (MT) technology (see [14] and references therein) of HTS bulks processing, a real interest in the levitation systems for so-called large scale (LS) applications [15, 16] has appeared. The LS systems like flywheels for energy storage, electric motors and generators, permanent magnets, etc. are very promising for HTS practical utilization and actively studied now in the whole world.

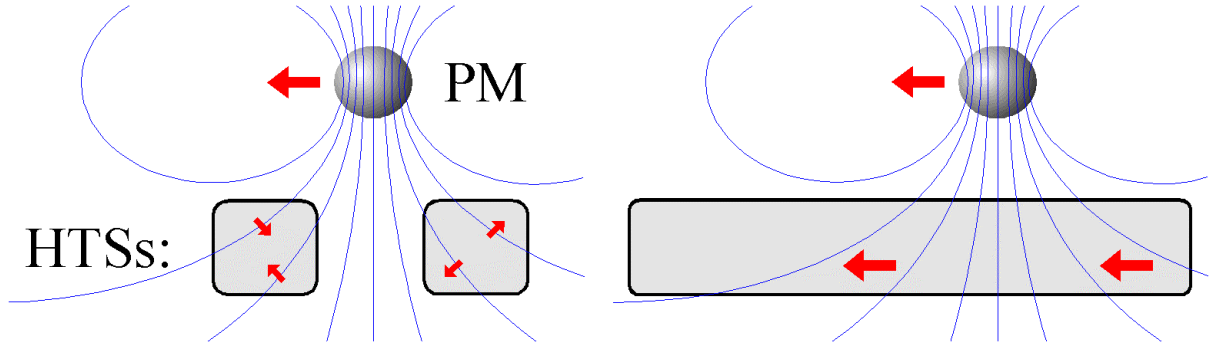

Fig. 1. Schematic illustration of vortex motion in superconducting samples (HTSs) for two extremes: LS << LF (left panel) and LS >> LF (right panel), where LS is a character sample size and LF is a scale of the external magnetic field variations. The permanent magnet (PM) is the source of the magnetic field and its motion causes the motion of the vortices in superconductors (shown by arrows). See text for details.

In the following sections we start with terminology explanation, then we describe some of the levitation techniques, their application to the investigation of compact vortex structures dynamics in HTS bulks, and, at the end, briefly discuss the practical aspect of the problem.

## 1. Terminology

We focus in this paper on the investigation of the macroscopic magnetic properties of HTS bulks. The word 'macroscopic' here designates the scale from the magnetic field penetration depth to a size of the sample. The notion 'bulk' implies that all sample's dimensions $L_S$ are comparable or bigger than the scale of the external magnetic field variations $L_F$. Fig. 1 illustrates the principal difference in vortex dynamics for two extremes: $L_S \ll L_F$ (left panel) and $L_S \gg L_F$ (right panel). For the sake of simplicity we consider an ideal type-II superconductor without pinning and surface barrier. A permanent magnet (PM) is shown as the field source and its motion produces a time changing field.

When the sample is much less than the field variation scale (much less than it is shown at the left panel), it treats the field as uniform ($\mathbf{H}(\mathbf{r}) = \mathbf{H}_a$), and evidently the vortex motion inside the superconductors do not coincide with the PM motion but is determined by this field value $\mathbf{H}_a$, its changing rate $d\mathbf{H}_a/dt$ and the size of the sample. We want to note here, that in the schematic draw in Fig.1 we neglect field distortions due to screening currents and show vortices as 'connected' to the magnetic lines, but in real geometries when a magnetic field configuration is changing with time, the concept of the 'magnetic lines' is useless and even wrong. In the opposite case (see right panel of Fig.1), when $L_S \gg L_F$, the motion of vortex bunch inside the sample simply follows to the PM motion. Samples of such a scale we call 'bulk'. The system shown at the right panel in Fig.1 is a simple extreme for which some analytical solutions of field configurations have been found. These models and the related experimental procedures are discussed in the following chapters. For the 'bulk' samples of an intermediate range ($L_S \sim L_F$), the field configurations remain mostly unresolved yet. In order to calculate the behavior of such systems, new empiric approaches should be introduced.

## 2. Techniques description

One can say that the development of LS HTS applications which include a number of systems based on superconducting levitation revealed a gap in our understanding of the macroscopic magnetic properties of the bulk superconductors and that the usual experimental techniques were mostly useless to study the problem, but fortunately, the levitation systems themselves had appeared to be able to do the job. Here we focus on the main levitation techniques which are widely used now to characterize the HTS samples and investigate the problem. They are levitation force measurements, the resonance oscillation technique, and the methods based on contactless magnetic rotors.

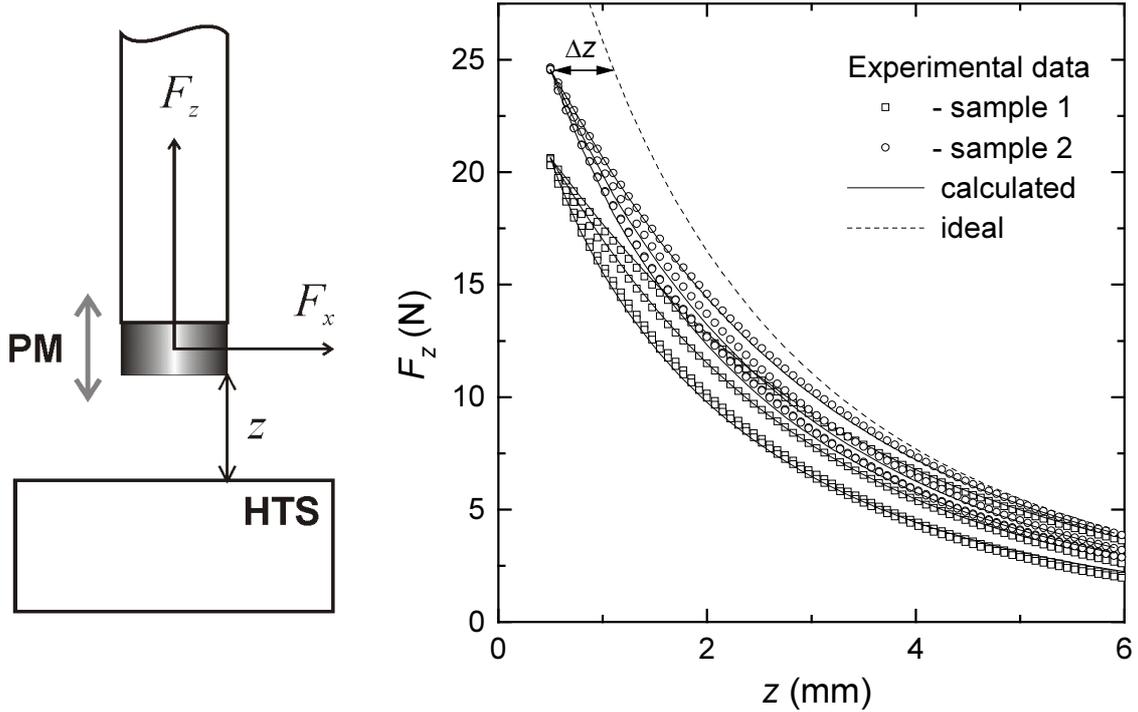

Fig. 2. Experimental configuration for levitation force measurements (left panel) and an example of the experimental data (right panel) [19].

These methods are non-destructive by nature, that gives an opportunity to carry out the different experiments with the same samples and obtain an information about vortex dynamics in HTS bulks in a wide range of frequencies and field amplitudes. We also briefly mention a couple of phenomena in PM-HTS interaction which have not been widely utilized yet but also enable to obtain unique results on vortex dynamics. The viscous PM motion through a HTS sample aperture is such an example.

2.1. LEVITATION FORCE MEASUREMENTS

Measuring the repulsion force between a piece of magnet and a superconductor is usually considered as a simplest way to detect the onset of the bulk superconducting state in the sample. This method had been known since the discovery of the Meisner effect but, being contactless in its nature, become extremely popular with the beginning of the HTS era only. In that time, the levitation force measurements were widely utilized in the processes of searching new superconducting compounds and developing new synthesizing techniques. Rapid development of the levitation applications mentioned above has led to further development of these measuring technique and now the levitation force measurements are usually considered in two roles: as an information source to know more about levitation systems and as a quick technique to test HTS samples [17, 18].

In general case, within a 'levitation force measuring procedure', one can arbitrarily move/rotate a permanent magnet (PM) of an arbitrary shape measuring the force and torque acting on magnet but, in order to simplify the data analysis, the symmetric configuration is more preferable. Fig. 2 (left panel) shows a sketch of such a simplified standard levitation force measuring procedure. In such an axially symmetric configuration, a PM moves up and down above the investigated HTS sample and the measuring value is the vertical levitation force $F_z$ as a function of distance $z$. Fig. 2 (right panel) represents an example of such experimental data. A lateral force, $F_x$ and/or $F_y$, if appears, indicates that the investigated sample is not symmetric, usually multidomain. The scale of the 'interaction volume' in the HTS sample, integrated properties of which one may decipher from such measurements, is determined by PM size and distance $z$ and, consequently, can be set by experimentalist. Using such a procedure, we have developed a simple approach to determine the critical current density in melt-processed HTS bulks [19]. It has been shown that, according to the simplest 'visual' method, the bulk critical current density of the sample can be estimated as

$$J_c \approx \frac{c}{8\pi} \frac{b_{ar}}{\Delta z}, \qquad (1)$$

where $b_{ar}$ is a parallel to the top HTS surface component of the PM field (its maximum value at level of the surface) and $\Delta z$ is the shift of the experimental $F_z(z)$ curve with respect to the calculated ideal one (see Fig. 2). A brief description of the approach is given in section 4.2 (for more details see [20]). Another approach [21, 22] has been developed to estimate the quality of superconducting joints (welds) between blocks of HTS bulks [23] and evaluate the joint's critical current density.

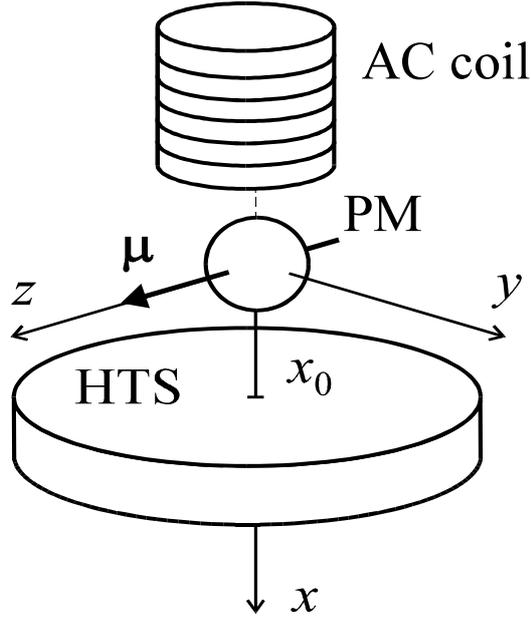

Fig. 3. A basic configuration of the resonance oscillations technique.

## 2.2. RESONANCE OSCILLATION TECHNIQUE

The resonance oscillations technique is simple to realize but allows to obtain the unique results in understanding of the magnetic properties of HTS bulks. This method is based on an investigation of a mechanical system with a freely levitated permanent magnet above a HTS sample. Measuring the parameters of forced oscillations of a PM one can obtain information about structure and electromagnetic properties of explored HTS samples. Fig. 3 represents a basic system [24–26] which has been used with granular [27–32] and composite [30, 32] YBCO and BSCCO ceramics, and, in a slight modification, with the melt-processed YBCO quasi-single crystals [11, 32–36].

Here, the forced oscillations of the permanent magnet with mass $m$, magnetic moment $\mu$, and diameter $d$, which levitates on the distance of $x_0$ over the surface of the HTS sample have been induced by the AC coil and their amplitude has been measured by the microscope with a 5 μm accuracy. To increase the accuracy, the magnetic induction system has been used. The PM produces the AC magnetic field $\mathbf{H} = \mathbf{H}_0 + \mathbf{h}_0 \sin(\omega t)$ at the surface and in the volume of the HTS. For the basic system on granular HTS the following values have been chosen: $m = 0.021$ g, $\mu = 1.2$ G cm$^3$, $d = 1.2$ mm, $x_0 = 2.3$ mm.

The main parameters here to measure are 'elasticity' and 'damping' (or 'energy loss'). It is possible to say that elasticity of the system is mainly determined by sample structure while the damping is mostly related with the energy loss which occurs due to magnetic flux motion in the HTS sample. In the given system, the 'elasticity' can be associated with the resonance frequency $\omega_0$ while the 'dumping' — with the dumping coefficient $\delta$ ($2\delta$ is the resonance curve width on a height of $A_{max}/\sqrt{2}$) or inversed $Q$-factor $Q = \omega_0/2\delta$. The storage energy $W_0$ and energy loss per period $W$ can be determined as:

$$W_0 = 1/2 m\omega_0^2 A^2, \quad W = 2\pi W_0/Q. \tag{2}$$

The dependences of these parameters on the resonance amplitude $A_{max}$ give a vast of information about structure and properties of HTS bulks but this information should be decipher first.

In the systems where it was possible to approximate the PM as a point magnetic dipole (with the $\mu$ parallel to the HTS surface, as it is shown in Fig. 3) there are five modes $s$ of PM oscillations: three translation modes, $s = x, y, z$ (along the corresponding axes); and two rotation (or torsion) modes, $s = \psi$ and $\theta$ (around $x$ and $y$ axes respectively) [31]. Every mode has the following parameters of its own: resonance frequency $\omega$, damping $\delta$, and their dependencies on the PM amplitude *A*. The presence of these modes highly enlarges the capabilities of the technique.

There are many similar configurations of the PM–HTS oscillating systems which have been studied by other authors [37–41]. We mention here three the most sensitive ones. (1) A superconducting suspension of the Nb sphere oscillating in the magnetic field has been used by Hebard [42] as an extremely sensitive tool in the experimental attempt to measure the quarks charges. (2) A tiny PM inside a superconducting capacitor has been used by Grosser et al. [43, 44] to study the vortex dynamics in HTS films. (3) The well known 'vibrating reed' method [45, 46] which, strictly speaking, does not belongs to 'levitation' techniques but is based on the same idea of measuring the elasticity and damping in a superconductor-magnet system has yielded a lot of interesting results about pinning and vortex dynamics in superconducting films [47, 48].

## 2.3. MAGNETIC ROTOR

As it was written before, investigating the energy loss in HTS samples in AC field is an important tool to study the mechanism of the magnetic flux motion in superconductors, and the resonance oscillating technique has turned to be very successful for non-destructing investigations of HTS bulks. But using this technique one can mainly study the energy loss as a function of the magnetic field amplitude — only small frequency range can be covered by changing the PM mass [27, 28] or considering the different modes of the PM oscillations [29, 32].

The magnetic rotor technique allows to measure energy losses in a wide frequency range [49–51]. This method is based on a high speed magnetic rotor on contactless HTS bearings [52, 53]. Due to low energy consumption, this method allows to obtain information in the frequency range of 30–3000 Hz [49, 50].

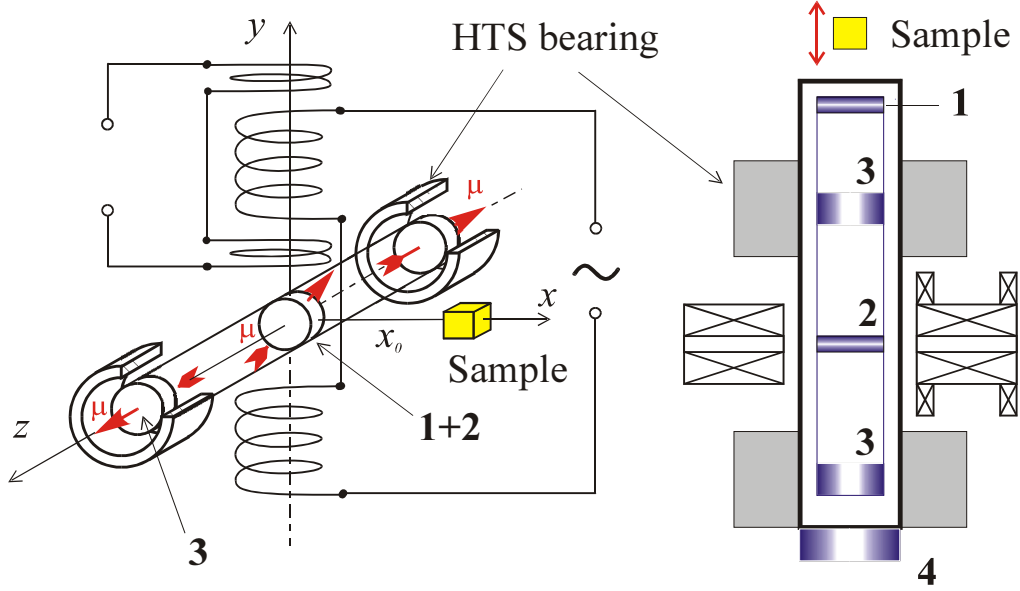

Fig. 4. Horizontal (left) and vertical (right) configurations of the magnetic rotor technique for the energy loss measuring: *1* is the magnet which rotates with rotor and generates the AC magnetic field in the sample, *2* is the driving magnet by which the rotor rotates, *3* are the carrying magnets which are the rotating parts of contactless bearings whereas the HTS parts play the role of a stator, *4* is a supporting magnet in the vertical configuration.

Fig. 4 represents the horizontal (left) and vertical (right) configurations of such experiments.

In the experimental procedures we have used either the free spin down measurements [51, 54, 55] or phase difference detecting [50–52]. The first technique was used by Hull et al. [54–56] in the investigations of the flywheel behavior to determine the coefficient of friction (COF). Exploiting such a procedure, for the energy loss we can write [51]:

$$W(\omega) = -2\pi J \frac{d\omega}{dt}, \qquad (3)$$

where $J$ is the inertial moment of the rotor, $\omega$ is its rotation frequency, and $d\omega/dt$ can be readily obtained from the experimental $\omega(t)$ dependency.

Another method of energy loss determination is the measuring of the phase $\varphi_0$ of the rotor rotation, which can be measured by phase difference between the driving and detecting coils (see Fig. 4). In this case the AC loss is [50, 51]

$$W(\omega) = \pi\mu\, H_0 \sin\varphi_0(\omega), \qquad (4)$$

where $\mu$ is the magnetic moment of the central driving magnet and $H_0$ is the AC field amplitude produced by the driving coils on this magnet.

Apart from these, there are different dynamic parameters of the rotor such as frequency and damping of the phase $\varphi_0$ oscillations, etc. that are determined by energy loss in the system and can be used for its measuring (see [51] for details).

## 2.4. OTHER TECHNIQUES

Both the resonance oscillations and magnetic rotor techniques have a low frequency limit about 10 Hz. Unfortunately, traditional measurements of the magnetic relaxation are limited, as a rule, by a 10 sec time window [57] which leaves the frequency range in 0.1–10 Hz uncovered by any technique. Thus, to fill the gap, new experimental techniques are highly wanted here, and the levitation techniques appeared to be suitable for this purpose too. Here, we briefly mention two of them.

One of such a technique is based on the effect of the viscous motion of a permanent magnet through the HTS aperture [32]. We have used it to investigate the magnetic flux relaxation with characteristic time about 1 sec (the short time magnetic relaxation) in granular HTS [58, 59]. Fig. 5 shows the experimental configuration which have been used [59].

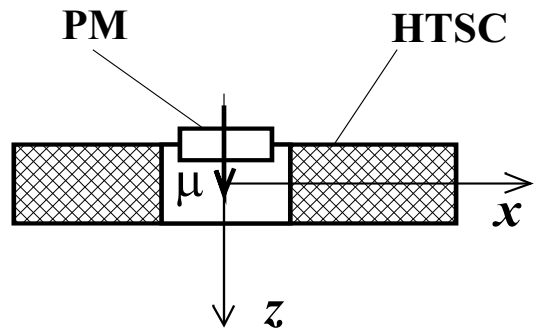

Fig. 5. The experimental configuration of the viscous motion method which have been used to measure short time magnetic relaxation in granular HTS [58, 59].

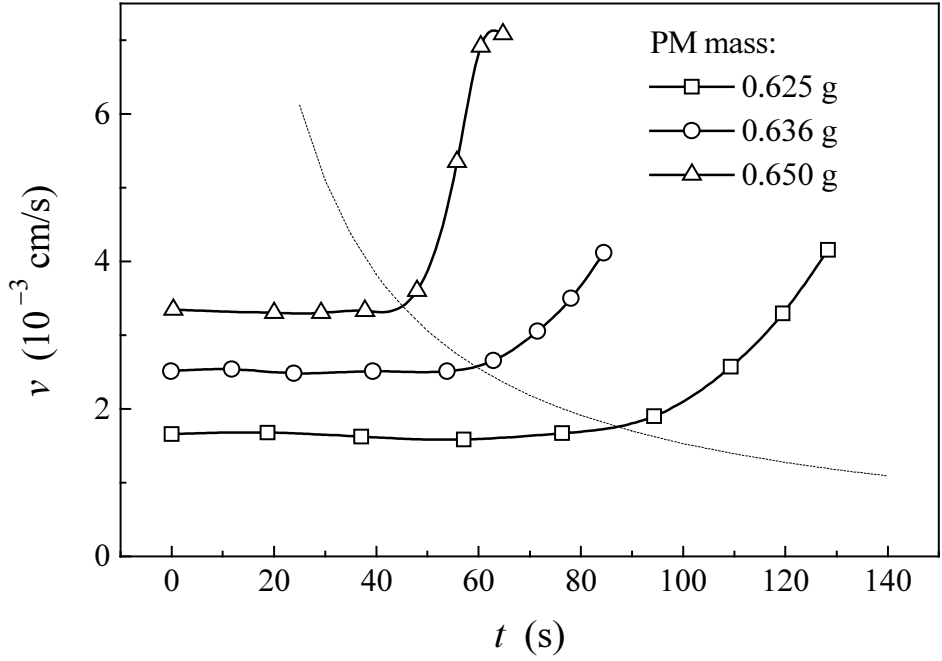

Fig. 6. Experimental dependencies of the PM velocity on time for different PM masses for BSCCO granular sample at 77 K. Dashed line shows the region where the PM begins "to feel" the HTS bottom edge.

During the experimental procedure, the magnetic moment of the PM was kept constant while the PM mass was varied by adding to the PM some of non-magnetic material. At 77 K under some relation of the experimental geometry, PM mass and magnetic moment, one can observe the viscous motion of the PM through the aperture of the superconducting sample under gravity. Fig. 6 represents the velocity dependencies of the PM on time: $v(t)$. The plateau, $v(t) = \text{const}$, allows us to relate this velocity with characteristic time of the magnetic flux diffusion in HTS grains. In [58, 59] we have obtained the relation

$$v = \frac{g}{\omega_z^2 \tau_0}\left(1 - \frac{m_0}{m}\right), \qquad (5)$$

where $m$ is the PM mass; $m_0$ is a maximum mass at which the magnet motion velocity is zero that was obtained from the data extrapolation for the PM with the different masses; $g$ is the gravity acceleration; $\omega_z^2$ is the resonance frequency of the PM vertical oscillations. The method allows to measure the characteristic time of the magnetic flux relaxation with the accuracy up to 5% [59].

Another interesting technique to study low frequency magnetic properties of HTS bulks has been proposed by Rossman and Budnick [60]. The authors explore both the experimental and theoretical aspects of the precession of a magnet over different HTS samples and have shown that the precession measurement can be utilized as a useful method to study low frequency (< 0.1 Hz) AC screening response loss of the superconductor.

In the following sections we present the results which have been obtained by applying a set of the levitation techniques described above.

## 3. Granular HTS

Historically, the first HTS-samples were the granular ceramics. It is these samples that have been exposed in the first place to the research boom on the beginning of the HTS era. And from the very beginning two main approaches have been used to describe the macroscopic magnetic dynamics in these samples: (i) as a motion of vortices (Abrikosov-like) [61] and (ii) considering the sample as an array of Josephson junctions (weak links) [62, 63]. Now one can say that these approaches are no longer conflicting. The question is not which one is better but where and when they should be applied. Evidently, in order to understand transport properties of granular HTS or their low field high frequency response the influence of intergrain junctions cannot be neglected. But low frequency magnetic properties of such samples in magnetic fields higher that 100 Oe are *mainly* related with intragrain vortex dynamics [30–32]. These properties are the magnetization, its relaxation and energy loss. And it is the interaction of superconducting grains with magnetic field that determines these properties and, consequently, the mechanical properties of large scale levitation system. This issue is addressed below in the following subsections.

We have investigated the $YBa_2Cu_3O_{7-\delta}$ (Y-123), $Pb(Bi)_2Sr_2Ca_2Cu_3O_{10+\delta}$ (Bi-2223) and $Pb(Bi)_2Sr_2CaCu_2O_{8+\delta}$ (Bi-2212) ceramics with typical size of grains about 10–20 μm for Y-123 and with 5–10 μm for Bi-compounds. Then, two structural types of samples of each compound have been used: the ceramic and the composite samples. All samples were 0.8 cm thick pellets of 4 cm diameter. The ceramic samples have been cut from the sintered superconductors. The composite samples have been formed from the HTS grains dispersed into insulating paraffin. The dispersed grains have been obtained by grinding the sintered superconductors. The absence of electrical contacts between grains in the composite samples has been checked measuring the resistance.

### 3.1. ROLE OF GRAINS

We start from the elastic properties of the PM-HTS systems. In [31] we have shown that, in order to describe the elasticity of such systems with Y-123 and Bi-2223 ceramics at 77K, the granular sample can be considered as a set of independent isolated grains. For the resonance frequencies of the PM $\omega_s$, for each of five modes of its oscillations, one can write an expression

TABLE 1. The geometric coefficients (see Eq. 6) for the PM-HTS oscillating system where the point magnetic dipole freely levitates above a granular HTS [29].

| Modes $s$ | $\zeta_s$ | $\xi_s$ | $\xi_s^{\parallel}$ | $\xi_s^{\perp}$ | $W_{\parallel}/W$, % | $W_{\perp}/W$, % |
|---|---|---|---|---|---|---|
| $x$ | 3 / 16 | 3 / 16 | 193 / 1134 | 157 / 9072 | 91 | 9 |
| $y$ | −3 / 64 | 3 / 64 | 193 / 6804 | 2015 / 108864 | 61 | 39 |
| $z$ | −9 / 64 | 9 / 64 | 383 / 24300 | 48547 / 388800 | 11 | 89 |
| $\psi$ | −1 / 16 | 1 / 16 | 29 / 6480 | 47 / 810 | 7 | 93 |
| $\theta$ | −1 / 16 | 1 / 8 | 29 / 1620 | 347 / 3240 | 14 | 86 |

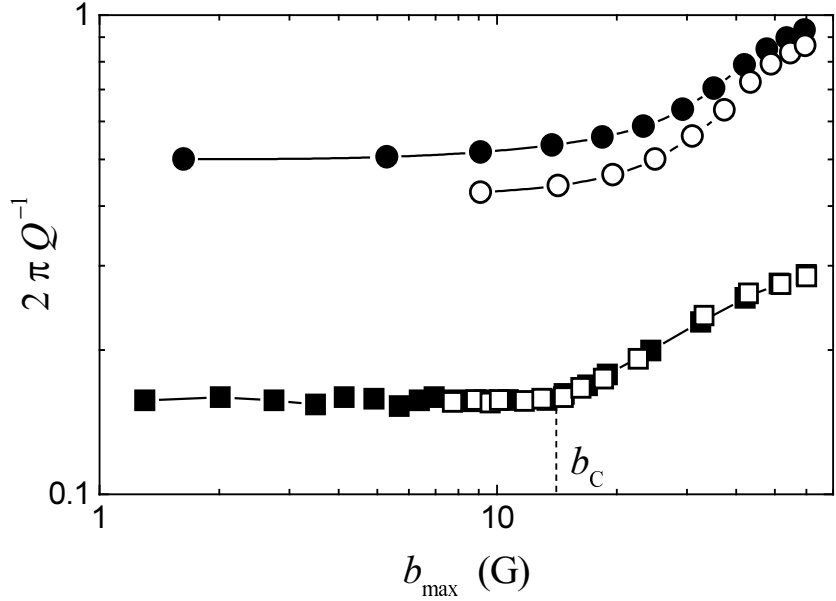

Fig. 7. The inverse $Q$-factor *vs* the maximal amplitude of magnetic field on the HTS surface for the ceramic (closed symbols) and composite (open symbols) samples made of Y-123 (squares) and Bi-2223 (circles) superconductors.

$$G_s \omega_s^2 = \frac{\alpha}{1-D}(\xi_s + \beta\zeta_s)\frac{\mu^2}{x_0^5}, \qquad (6)$$

where $G_s$ is the generalized mass: $G_x = G_y = G_z = m$, $G_\psi = G_\theta = (1/10)m(d/x_0)^2$, $\xi_s$ and $\zeta_s$ are the geometrical coefficients represented in Table 1, $\alpha$ is the superconducting grains volume fraction, $\beta$ is a volume magnetization coefficient (see [31] for details), and $D$ is a demagnetization factor which takes into account the field of grains: the effective magnetic field for each grain is $\mathbf{H}_i = (\mathbf{H}_0 - D\mathbf{B})/(1-D)$, where $\mathbf{H}_0$ is the field of the PM and $\mathbf{B}$ is the 'magnetic induction' or the field inside grains. The demagnetization factor is very close to 1/2 for the 'dense' granular systems with $\alpha > 0.5$ [13]. Thus, the 'physical' coefficients $\alpha$ and $\beta$ can be determined from the experimental values of the resonance frequencies for any two modes.

In [30], comparing the ceramic and composite samples, we have shown that, from point of view of the AC loss, the Y- and Bi- superconducting ceramics also can be considered as a set of independent grains. Fig. 7 represents the inverse $Q$-factor as a function of $b_{max}$, the maximum field amplitude on the HTS surface for the granular and composite samples. The $Q$-factor of this system is independent on $x_0$, $\alpha$ and $A$, so it is convenient to compare the different samples.

## 3.2. VISCOSITY EVALUATION

The plateau, that can be found on the dependencies of $Q^{-1}$ (see Fig. 7) on the amplitude, shows the viscous mechanism of the energy losses for which [24, 30]

$$W = 2\pi m \delta \omega_0 A^2 \sim b_{max}^2. \qquad (7)$$

Above the critical amplitude $b_c \sim 15$ G, the hysteretic loss occurs [27].

Then, using simple reasoning [30], one can evaluate the range of the viscosity of the intragrain flux motion. The linear viscosity $\eta_l$ is the proportionality factor between the viscous frictional force per vortex length and the vortex velocity: $\mathbf{f}_v = -\eta_l \mathbf{v}$. One can write the volume viscosity $\eta_V = \eta_l B / \phi_0$, where $B$ is the magnetic field in the superconducting grains and $\phi_0$ is the magnetic flux quantum. Than the energy loss (7) can be rewritten as

$$W = \pi \eta_l \omega \langle a^2 \rangle V_0 B / \phi_0, \qquad (8)$$

where $\langle a^2 \rangle$ is the mean square oscillation amplitude of vortices and $V_0$ is the volume of the superconductor in which the main energy dissipation occurs. For a quantitative estimation of $\eta_l$ one should know the mechanism of the vortex penetration into grain but it is possible to estimate the lower limit of $\eta_l$. In Ref. 30 we have obtained $\eta_l > 3 \cdot 10^{-4}$ g/cm sec (or $\eta_V > 10^{-5}$ g/cm$^3$ s) that exceeds the flux flow viscosity $\eta_{FF} \sim 10^{-7}$ g/cm sec [24] in more then three orders of magnitude. Such a giant viscosity may be explained by another linear mechanism of the viscous flux motion, namely the thermally assisted flux flow (TAFF) [64, 65].

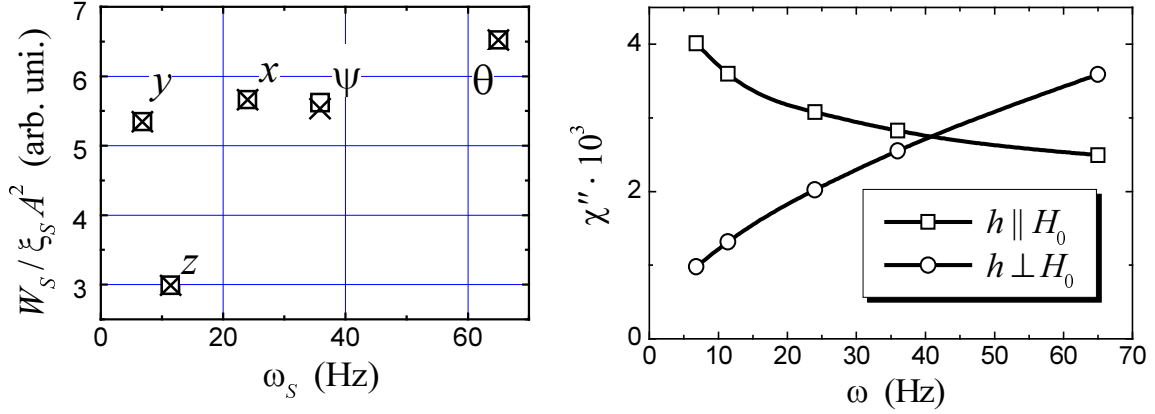

Fig. 8. Experimental (□) and calculated (×) values of AC loss per square amplitude *vs* resonance frequency for five modes (left panel), and the recalculated imaginary part of the susceptibility *vs* alternating magnetic field frequency for longitudinal and transverse field components for Y-123 granular sample (right panel).

## 3.3. TRANSLATIONAL AND ROTATIONAL FIELD COMPONENTS

For the viscous flux motion the imaginary part of the magnetic susceptibility $\chi''$ does not depend on AC field amplitude $h_0$ but only on its frequency. The energy loss per period can be obtained by taking integral over all HTS sample volume $V$:

$$W = \pi\alpha\chi'' \int_V h_0^2(\mathbf{r}) d\mathbf{r}^3 . \qquad (9)$$

So, due to viscous character of the AC losses under $b_c$ (7) in the PM-HTS system for each mode $s$ ($\mathbf{h}_s = (\partial \mathbf{H}/\partial s) A$) one can write [27, 28]:

$$W_s = 4\pi^2 \alpha \mu^2 x_0^{-5} \xi_s \chi''(\omega_s) A^2. \qquad (10)$$

Here, for the rotational modes: $A = \varphi/x_0$, where $\varphi$ is the angular amplitude of the PM oscillations. The numeric values of $\xi_s$ are the same as in Eq. 6 and also presented in Table 1. The non-monotonic dependence of $\chi'' \propto W_s/A^2\xi_s$ on $\omega$ which is shown on the left panel of Fig. 8 demonstrates that it is not possible to explain the data within an isotropic model, neglecting AC field direction. So, as a first possible step, we separate the contributions of the translational $h_\parallel$ (parallel to DC field $\mathbf{h}_\parallel \| \mathbf{H}_0$) and rotational $h_\perp$ (perpendicular, $\mathbf{h}_\perp \perp \mathbf{H}_0$) AC field components. Then $h^2 = h_\parallel^2 + h_\perp^2$ and $W = W_\parallel + W_\perp$. The numeric values of $\xi_S^\parallel$ and $\xi_S^\perp$ are also presented in Table 1. The relative contribution from this components to AC loss is different from mode to mode (see Table 1). For the $x$ and $y$ modes the translational component predominates; for other tree modes: $z$, $\psi$ and $\theta$, this is vice versa.

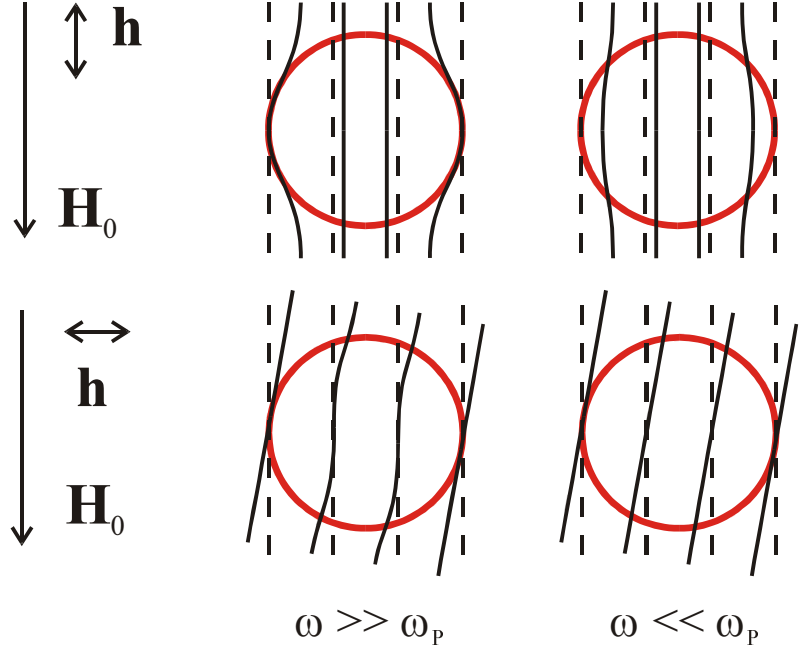

Fig. 9. A schematic representation of the penetration of the parallel and perpendicular AC field components into a superconducting grain.

Now, from the experimental values of $W_s$ using numerical calculations, one can find $\chi_\parallel''(\omega)$ and $\chi_\perp''(\omega)$ functions. The result of such calculations for the YBCO sample is shown in Fig. 8 (right panel). The calculations were made for the four modes and checked for the fifth one (see Fig. 8). These data were also verified in the direct measurements where the resonance frequency was changed by changing the PM mass [27, 29].

The reason for the differences in the frequency dependencies between the translational and rotational components becomes obvious if one remember that the translational component tends to change the magnitude of the intragrain magnetic flux by trying to penetrate through the grain boundary but the rotational component causes the change in the intragrain magnetic flux direction only (see Fig. 9). It means that the differences between $\chi''_\parallel(\omega)$ and $\chi''_\perp(\omega)$ functions can be explained by the grain surface barrier for the magnetic flux to penetrate into grains [66, 67].

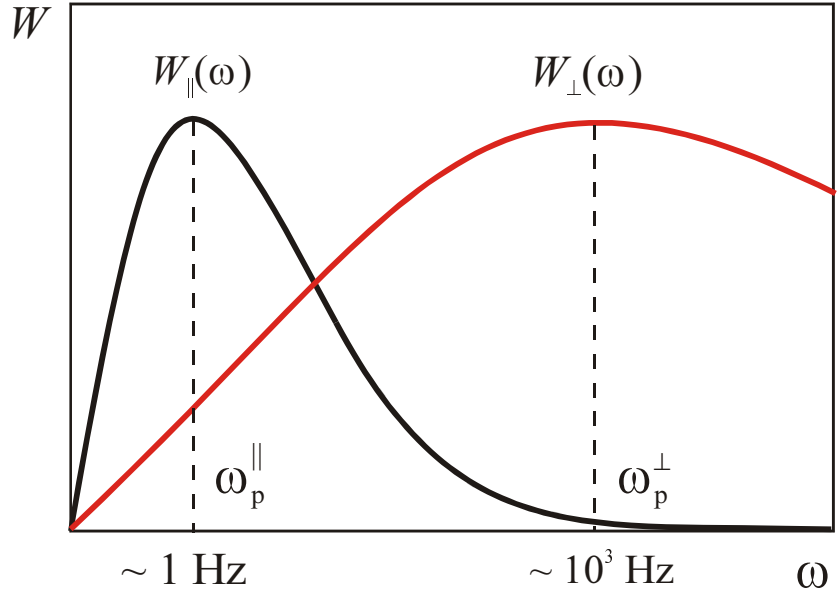

Fig. 10. Schematic frequency dependencies of the AC loss for the parallel and perpendicular AC field components.

## 3.4. VISCOUS ENERGY LOSSES

The results presented above prove the viscous nature of the energy loss in granular HTS for the amplitudes less then 15 G for the vortices motion both in the grains volume and through the grains surface. It was shown that the energy losses are induced by TAFF with very high viscosity, but to determine this viscosity it is necessary to find a maximum on an energy loss frequency dependence.

To investigate the dependence of $W_\perp$ on frequency the magnetic rotor technique has been used [49, 50]. In its horizontal configuration (left panel of Fig. 4), the rotor induces the superposition of the rotational and translational

AC magnetic field components at the surface of an explored sample. But over ~ 100 Hz rotational speed, the contribution of the translational component is negligible and it is possible to consider that all losses in HTS are induced by the viscous vortex motion in the volume of grains. The obtained numeric value $\eta_l =$ 8·$10^{-5}$ g/cm sec [49] coincides well with the above estimation. The characteristic time of the rotational field component penetration is $10^{-3} - 10^{-2}$ sec [32].

An additional information on the $W_\parallel(\omega)$ dependence can be obtained by the PM viscous motion method. Under velocities of about $10^{-3}$ cm/s the perpendicular component is negligible (see Fig. 8, right panel) and the slow viscous motion of the PM through the HTS aperture is mainly determined by the thermally activated diffusion of the magnetic flux into grains through the surface barrier [58]. From the characteristic time $\tau_0 = 1\pm0.07$ sec obtained from (5) [59] we can conclude that the viscosity of this process is by two-three orders of magnitude higher than that of in grains and $\eta_l \approx 10^{-2}$ g/cm sec. The $W_\parallel(\omega)$ and $W_\perp(\omega)$ dependencies are schematically represented in Fig. 10 [32].

In conclusion, we have shown that the interaction forces between a PM and granular HTS, the energy loss and relaxation in such systems are determined by magnetic field interaction with superconducting grains (at the nitrogen temperatures and DC field up to 200 G). The mechanisms of such interaction are following:

- The penetration of the parallel AC field component into Bi-2223 and Bi-2212 ceramics can be described by thermally activated diffusion of vortices through the surface barrier ($H_s$ = 5–8 Oe) with characteristic time about 1 sec.
- The vortex motion inside superconducting grains of Bi-ceramics exhibits a viscous nature, with $\eta_l \approx 5 \cdot 10^{-3}$ g/cm sec, and, below $h < h_c \sim 15$ Oe (for $\omega \sim$ 100 Hz), is determined by TAFF — in this field-frequency range the magnetization of grains can be well described by the viscous flux diffusion model and the maximum in $W(\omega)$ dependencies correspond to a 'full penetration' of AC flux into grains, depends on grain size $d$ and AC field amplitude $b$, $\omega_p = \phi_0 b/(2d^2\eta)$, and is about $10^3 - 10^4$ Hz. With increasing the field amplitude over $h_c(\omega)$ (or the frequency over $\omega_c(h)$), this diffusion becomes nonlinear by means of so called flux creep (FC) which then transforms into flux-flow (FF) with frequency independent energy loss.

- The nature of the flux dynamics inside Y-123 grains is the same but the parameters a different: $\eta_l \approx 5 \cdot 10^{-3}$ g/cm sec, $h_c \approx 22$ Oe, $\omega_c \approx 10^2 - 10^3$ Hz.
- The penetration of the parallel AC field component into Y-123 grains is also governed by the surface barrier but with twice bigger value, $H_s \approx 12$ Oe. Because of this, the thermal diffusion at ~ 80 K is much slower and at $h > H_s$, after quite narrow FC region, the surface related AC loss becomes hysteretic (frequency independent).

## 4. Melt-processed HTS

The melt-processed large grain HTS samples have appeared to be very different from the granular ones in the levitation properties. First, from point of view of different levitation systems they are very close to ideally hard superconductor — they have very strong pinning resulting in the absence of the effect of the PM rise above HTS sample (flux repulsion) during its cooling. Second, the small isolated grains approximation does not work for large grains [33].

4.1. ZERO APPRPOXIMATION: IDEALLY HARD SUPERCONDUCTOR

In [33–35] we have shown that the approach of ideally hard superconductor is very useful to describe the elastic properties of levitation systems with the melt-textured HTS and as a first approximation to calculate the AC losses. The idea of the approach is to use the surface shielding currents which screens the sample from the penetration of the magnetic field variation (in respect to its initial, frozen distribution over the surface) to calculate the magnetic field distribution outside the superconductor and to obtain from this the elastic properties of the PM–HTS system. The magnetic field **B**(**r**) inside such an ideal superconductor does not change with any PM displacements. The feasibility of this approximation is roughly determined by the condition $d \ll L$, where $d$ is the field penetration depth and $L$ is the character system dimension (mainly the distance between PM and HTS). For more detailed consideration of the feasibility range, see [68]. Within such an approximation this problem has an exact analytical solution in case of a magnetic dipole over a flat superconductor in the field cooled (FC) case.

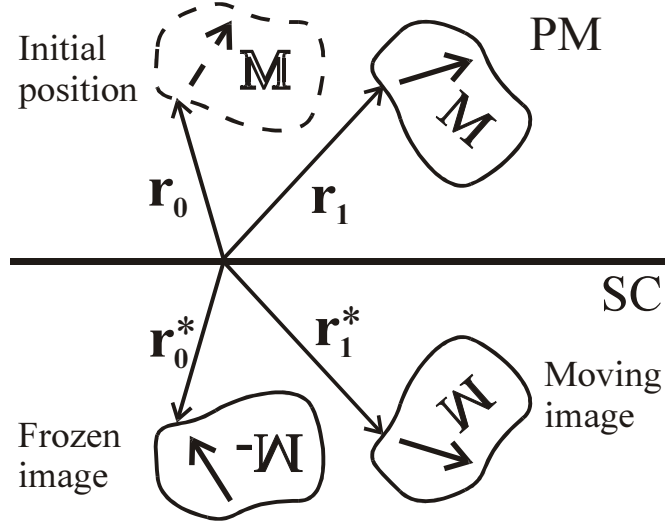

Fig. 11. Advanced mirror image method.

The nice illustration of this approach is the advanced mirror image method [33, 34]. Its distinction from the usual one, which is applied to the type-I superconductors, is in using of the frozen PM image that creates the same magnetic field distribution outside the HTS as the frozen magnetic flux does. Fig. 11 illustrates this. The configuration of the images shows that the ideally hard superconductor in FC case shields the magnetic field variation induced by magnet displacements. The rigorous prove of this method is given in [68].

Despite on such a simple illustration the analytical solutions can be obtained for the interaction of the point magnetic dipole with a superconductor for some of the simplest geometry such as a plane or an ellipsoid only. In any other cases one should use numerical calculations. Some analytical solutions for the levitation forces, resonance frequencies and non-linearity coefficients in the case of PM dipole over a flat superconductor, as well as their good agreement with the experiments, have been published in [33, 34, 68]. A perfect agreement of the frozen mirror image model with the experimental data has been also demonstrated with the levitation force measurements [69, 70].

4.2. FIRST APPROXIMATION: CRITICAL CURRENT DENSITY

Next step from the ideally hard superconductor approximation is from infinitely thin layer of undersurface shielding currents to the finite one which is thin enough (in respect to other system dimensions). In this case the hysteresis of the elastic properties and AC losses can be obtained.

In many earlier works (see [37, 71] for example) it has been shown that the forces between a PM and HTS sample are closely related with HTS magnetization curves. Vertical levitation force versus vertical distance $F_z(z)$ is the nearest analog to $M(H)$ dependencies with their major and minor hysteresis loops but the complexity of a field configuration in such large scale PM–HTS systems makes it very difficult to directly correlate them in general case. The problem can be solved by numerical approaches and some of them have been successfully used [72, 73]. The numerical approaches are undoubtedly useful to evaluate the real system parameters but usually need too much computer resources to be applicable to direct HTS sample investigation. To perform such an investigation an analytical evaluation is more wished for.

In [19, 20] we have shown that the first approximation mentioned above is already enough to describe the deviation of the experimentally measured levitation force from the ideal one, its hysteresis (see Fig. 2), and, consequently, to determine the critical current density $J_c$ from such measurements. Both 'visual' method presented in Fig. 2 and methods based on analytical calculation of the levitation force (see [19, 20]) utilize the idea that the magnetic flux penetration into an extremely hard superconductor forms a critical supercurrent layer under its surface, in respect to the critical state model [74]: if the sign of the external field variation is changed, a new critical supercurrent layer with an opposite direction starts to expand from the surface [19]. The depth of the layer

$$\delta = \frac{c}{4\pi}\frac{b(\rho)}{J_c}, \qquad (11)$$

where $c$ is speed of light and $b(\rho)$ is the AC field amplitude distribution over the sample surface (from the time it has started to change in given direction) which, and this is a condition of the first approximation, is twice of the parallel to the surface component of the PM field corresponding variation: $b(\rho) = 2\, b_{ar}(\rho)$ [20].

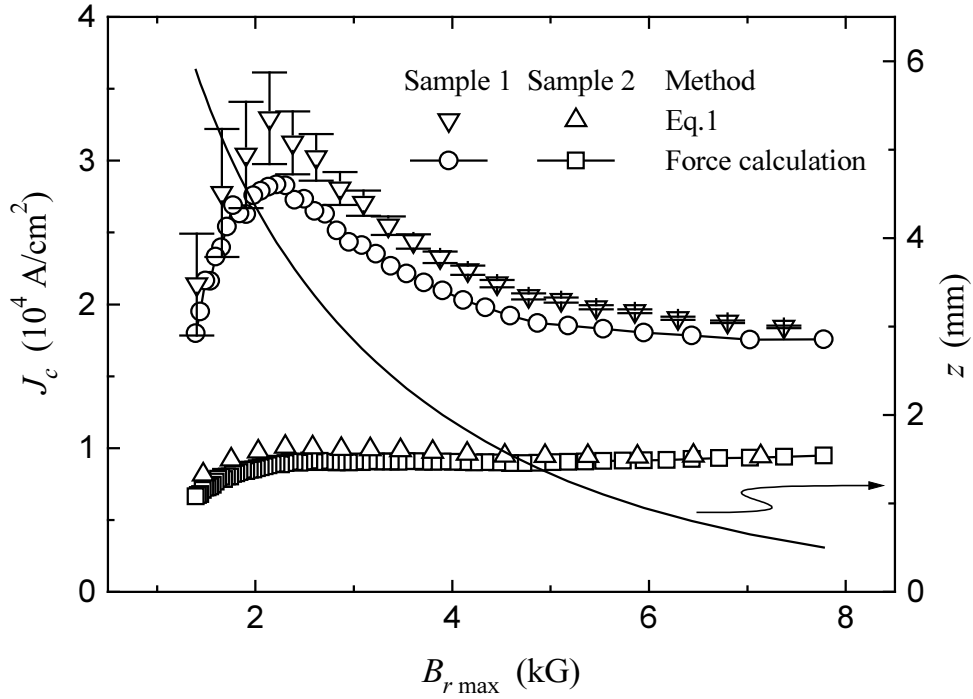

Fig. 12. The values of averaged critical current density versus maximum value of $B_r$, the magnetic field tangential component at the HTS surface, obtained by different methods [19]. The solid line with respect to right axis represents the dependence of $B_{r\max}(z)$.

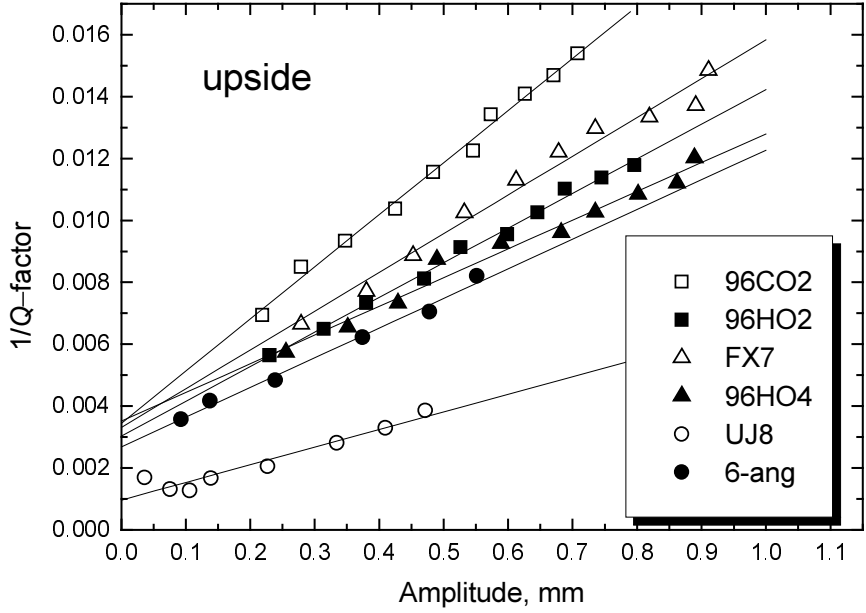

Fig. 13. Experimental data of inverse *Q*-factor *vs* PM amplitude for upper sides of different melt-processed HTS samples [35].

With this simple idea the calculated levitation force perfectly fits both the experimentally measured one and its hysteresises [19] (see Fig. 2). Fig. 12 shows the critical current densities obtained in such a way: the values of averaged critical current density versus maximum value of $B_r$, the magnetic field tangential component at the HTS surface, obtained by different methods. The solid line with respect to right axis represents the dependence of $B_{r\max}(z)$.

4.3. FIRST APPROXIMATION: ENERGY LOSSES

For the melt-processed HTS where the energy losses have predominantly hysteretic nature [34, 35] the feasibility of such an approximation can be determined from the critical state model (the thickness of the layer δ carrying the critical current $J_c$ must be well less than the PM–HTS distance: $\delta \ll z_0$, see Eq. 11). As far this is fulfilled, and if the dimension of the HTS sample is much more than $\delta$, the energy loss per square $w(\rho) = (c/24\pi^2)\, b^3(\rho)/J_c$. Integration of this energy loss over superconducting surface for vertical PM oscillations gives

$$W = \frac{2c}{3\pi}\frac{A^3}{J_c}\int_0^\infty \left(\frac{dB_r}{dz}\right)^3_{z=z_0} r\,dr, \qquad (12)$$

where $B_r$ is the parallel to the surface PM field component, and $z$ is vertical axis. From here, the inverse $Q$-factor of the PM—HTS system

$$Q^{-1}(A) = \frac{2c}{3\pi^2}\frac{A}{J_c m\omega_0^2}\int_0^\infty \left(\frac{dB_r}{dz}\right)^3_{z=z_0} r\,dr. \qquad (13)$$

The condition $\delta \ll z_0$ is much stronger than it is necessary to validate the use of the described approach for the melt-textured HTS. Even for $J_c \sim 10^4$ A/cm² and for $b \sim 100$ G the penetration depth $\delta \sim 0.1$ mm.

Fig. 13 presents the experimental data obtained with the slightly modified resonance oscillation technique [35]. The linear dependencies for the inverse $Q$-factor ($Q^{-1} \sim \alpha + \beta A$) mean that energy losses have two components:

$$W = \alpha A^2 + \beta A^3. \qquad (14)$$

The nature of the first component (viscous or surface losses [35]) is still unclear [13] but the second component is well known hysteretic losses that described by the critical state model. From this part using from Eq. 13 the value of the critical current density can be obtained.

The linear dependencies of $Q^{-1}(A)$ like represented in Fig. 13, are observed for the samples just after manufacturing. With aging of samples which firstly affects undersurface layers, these dependencies become non-linear [36]. This means that the critical current density $J_c$ becomes spatially nonuniform. Fig. 14 presents such dependencies for one melt-processed HTS sample just after manufacturing and half a year later. For the initial sample with the above approach we have obtained the value $J_c = 3.5\cdot10^4$ A/cm². The estimation of the undersurface $J_c(x \to 0)$ of the degraded sample from the initial slope of its $Q^{-1}(A)$ dependence gives: $J_c(0) = 0.79 \cdot 10^4$ A/cm². In [36] we have found the correlations between $J_c(x)$ and $Q^{-1}(A)$ functions and developed a technique to recover the critical current density profiles $J_c(x)$ in thin undersurface layer. The similar approach for reconstruction of critical current density profiles from AC susceptibility measurements have been proposed by Campbell [75].

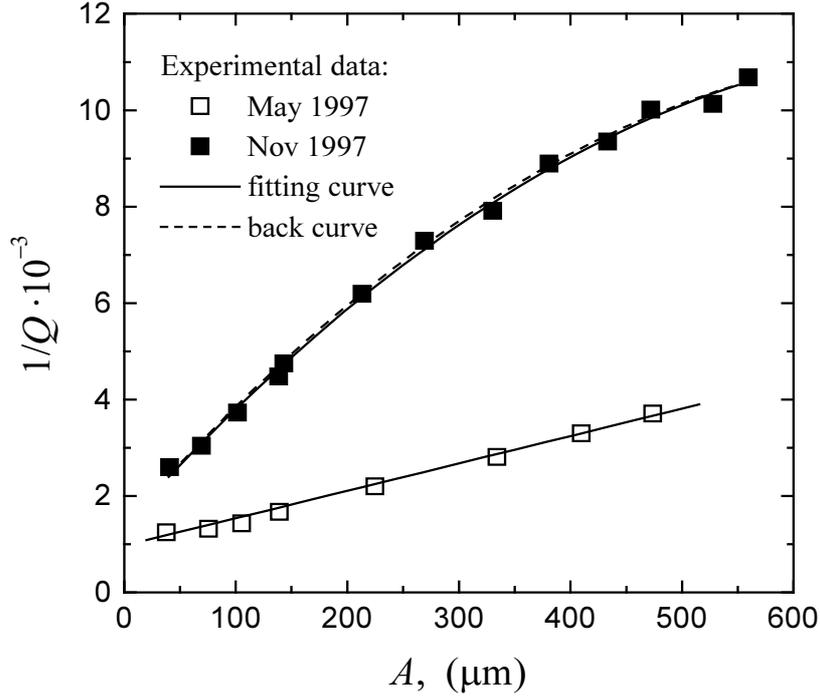

Fig. 14. Experimental dependencies of the inverse $Q$-factor on the resonance amplitude of the permanent magnet above the bulk YBCO sample just after manufacturing (□) and in half a year (■).

4.4. SECOND APPROXIMATION: CRITICAL CURRENT PROFILES

In [36] we have shown that in the AC magnetic field the energy dissipated per period per unit of surface area in non-uniform superconductor ($J_c(B) = $ const, $J_c(x) \neq$ const) is

$$W_s(b_s) = \frac{1}{\pi} \int_0^\delta b(x)\,(b_s - b(x))\,dx. \tag{15}$$

Here δ is the AC field penetration depth, $b_s$ is the amplitude of the AC field at the sample surface, and the AC field amplitude inside a superconductor is

$$b(x) = b_s - \frac{4\pi}{c}\int_0^x J_c(\xi)\,d\xi, \text{ for } 0 < x < \delta. \tag{16}$$

Then, from (15) and (16) one can readily obtain the critical current density $J_c$ at the depth δ as a function of the experimental data $W_s(b_s)$:

$$J_c(\delta(b_s)) = \frac{c}{4\pi^2}\left(\frac{d^2 W_s}{db_s^2}\right)^{-1} b_s. \tag{17}$$

Strictly speaking, Eq. 17 gives the dependence $J_c(b_s)$ at the depth $\delta(b_s)$ that provides some information about the real critical current profile but to obtain $J_c(x)$ itself in general case the numerical calculations should be applied. Below we consider an example when, in a specific case, it was possible to solve the problem analytically.

We analyze the experimental data presented in Fig. 14 which have been obtained by the resonance oscillation technique with slightly modified geometry: a permanent magnet (SmCo$_5$, ⌀6.3×2.3 mm, $m$ = 0.6 g, μ = 38 G cm³) levitates above the melt-textured HTS sample (⌀32×16 mm) with the magnetic moment perpendicular to the surface. We use a field cooled case with the initial distance between the permanent magnet and the HTS surface $x_0$ = 4.8 mm. The system is symmetrical about the vertical $x$-axis. A more detailed description of this configuration has been given in [35].

We consider the hysteretic part only that can be fitted for this sample by the polynomial $Q^{-1}(A) = q_1 A - q_2 A^2$, where $q_1$ = 0.26 cm$^{-1}$ and $q_2$ = 1.65 cm$^{-2}$.

The inverse $Q$-factor can be expressed as

$$Q^{-1}(A) = \frac{2}{m\omega^2 A^2}\int_0^R r\,W_s(b_s(r,A))\,dr, \tag{18}$$

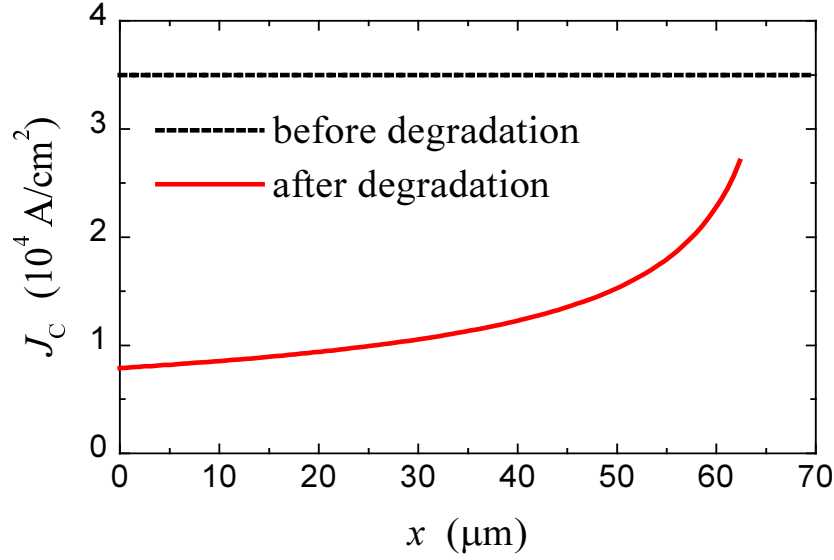

Fig. 15. Critical current density profiles inside the HTS sample before (dashed line) and after (solid line) degradation calculated from AC loss experimental data.

where $\omega$ is the resonance frequency of the PM vertical oscillations, that slightly depends on $A$: $\omega = \omega_0 - \gamma A$, $\omega_0 = 36.3$ Hz, $\gamma = 0.1$ Hz/cm, and $R$ is the radius of the HTS sample. For a flat extremely hard superconductor (see [34]) at its surface we can write $b_s(r,A) = 2\theta(r)A$, where $\theta(r) = dB_r(r)/dx$, and $B_r$ is the component of the PM field that is parallel to the surface of the sample. For simplicity we assume that $b_s = 2\theta_{\text{eff}} A$ and use the value $SW_s$ instead of the integral in (14). Then

$$W_s(b_s(A)) = \frac{\pi m \omega_0^2 b_s^2}{4\theta_{\text{eff}}^2 S} Q^{-1}(b_s(A)). \tag{19}$$

We used the empirical formula $S = 2\pi k R x_0$, and in our configuration, $\theta_{\text{eff}} = 0.87\,\theta_{\max} = 870$ G/cm, $k = 0.94$. Then, using this assumption we obtain $J_c(\delta(b_s)) = J_{c0}/(1-\varepsilon b_s)$ with $\varepsilon = q_2/(\theta q_1)$ and finally

$$J_c(x) = \frac{J_{c0}}{\sqrt{1-px}}, \tag{20}$$

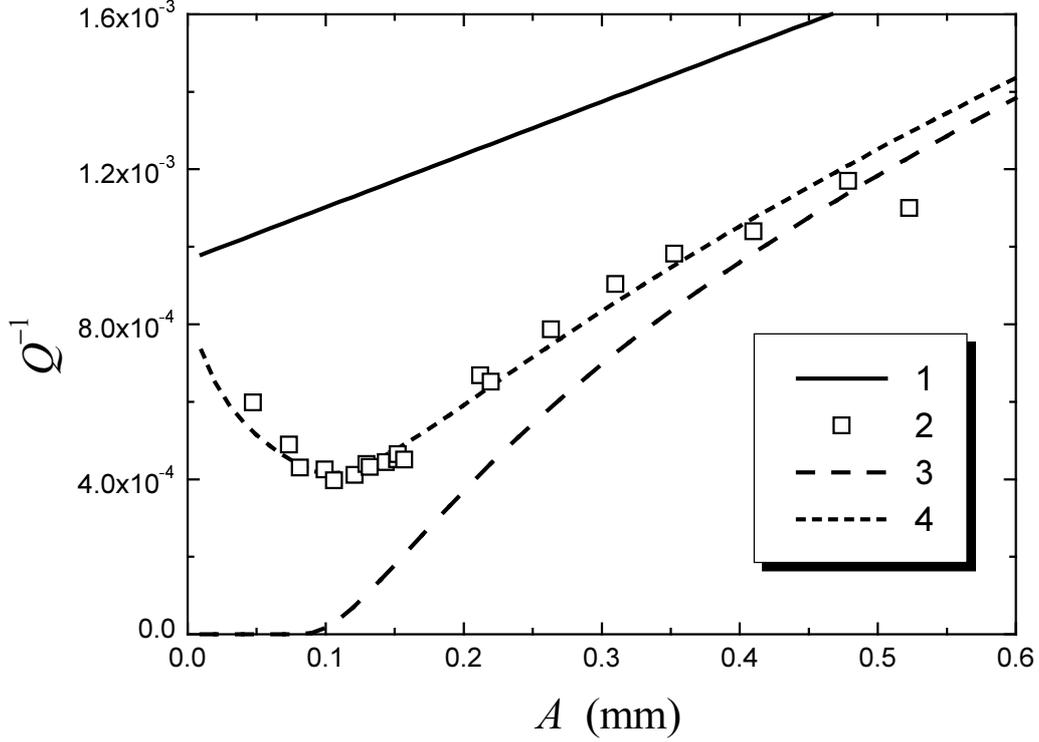

Fig. 16. Inverse $Q$-factor vs PM amplitude: experimental, before (*1*) and after polishing (*2*), and calculated data with the Bean-Livingston surface barrier (*3*) and for the model of the dynamic barrier surmounting (*4*).

where $J_{c0} = c\theta S/(3\pi^3 m\omega q_1) = 0.79 \cdot 10^4$ A/cm$^2$, and $p = 8\theta S q_2/(3\pi^2 m\omega q) = 147$ cm$^{-1}$. The function $J_c(x)$ is shown in Fig. 15.

The dashed curve ("back" curve) in Fig. 14 represents the dependence obtained back from (20), (16) and (15) with real integrating in (18). The values $\theta_{\text{eff}}$ and $k$ were chosen for this curve to coincide with experimental data in two "points": for the initial slope determined by $J_{c0}$, and for $Q^{-1}(A_{\max})$. The good coincidence of this back curve with the fitted curve on the hole range gives the proof of feasibility of Eq. (15) instead of (14).

## 4.5. VORTEX PENETRATION THROUGH THE SURFACE BARRIER

An interesting effect appears after the polishing of the surface of melt-processed HTS samples: the minimum at $Q^{-1}(A)$ dependencies [11] (see Fig. 16). We have shown that this effect can be explained in terms of a dynamic crossover from the absence of the surface barrier at low rates of field variation to its appearance at higher rates.

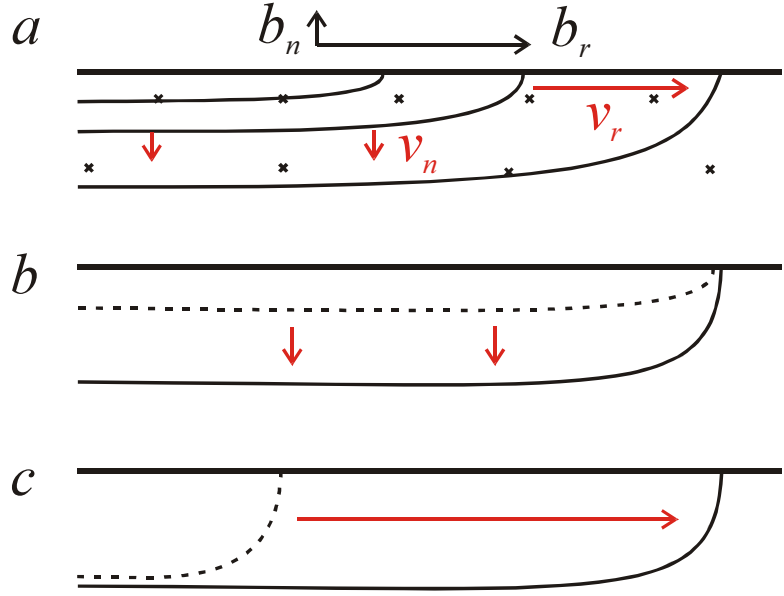

Fig. 17. Scheme of the compact vortex structure which is formed under HTS surface by the oscillating PM (*a*). Panels (*b*) and (*c*) represents two extremes of vortex propagation.

The idea of such a dynamic barrier is following. As a result of the strong pinning, the vortices are formed in long half-loops under HTS surface, as it is schematically shown in Fig. 17 (*a*). As far as the parallel to the surface AC field component is much bigger than the perpendicular one: $b_r \gg b_n$, the velocity, $v_r$, of the along surface propagation of perpendicular parts of vortices should be much higher than the velocity of parallel parts of vortices, $v_n$. Then one can consider two extremes for a vortex to reach its position (which is determined by $b_r(\rho)$ distribution): shown as (*b*) and (*c*) in Fig. 17. In presence of a surface barrier, the last one corresponds to energy-minimum motion for low $v_r$ but becomes unfavorable at high velocities. Thus, there is a critical velocity $v_{rc}$ at which an influence of the surface barrier on compact vortex structures penetration appears. The calculations (dotted line *4* in Fig. 16) which based on this idea and on consideration of the Bean-Livingston surface barrier [66, 76] gives very good agreement with the experiment [11, 13].

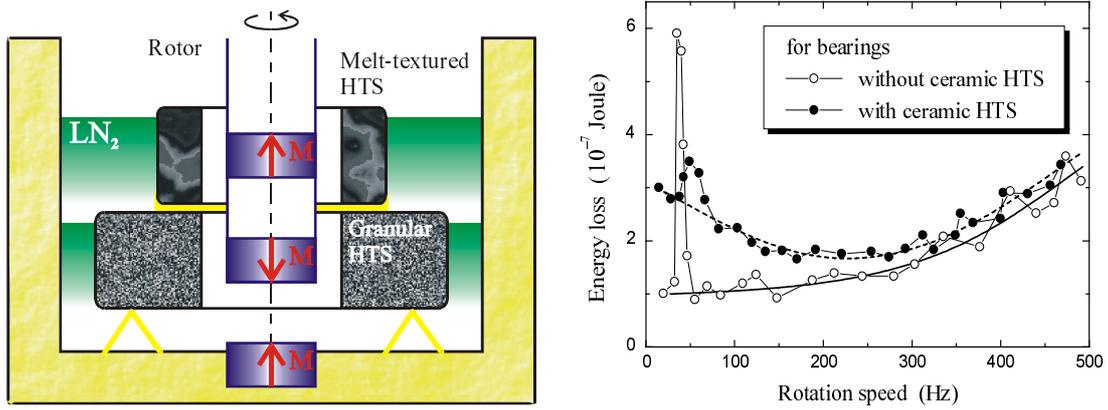

Fig. 18. Scheme of the compound bearing (left) and energy loss in such a system with and without the granular component (right).

To summarize, we have shown that the vortex dynamics in melt-processed quasi-single crystals of Y-123 has essentially hysteretic nature — the low frequency energy loss does not depend on AC field frequency — and can be well described by the critical state model. The whole physics of the PM–HTS interaction is concentrated in a thin undersurface layer of HTS samples where the compact vortex structures alternately penetrates into HTS during PM oscillations. Consequently, the dynamic parameters of the PM–HTS systems are very sensitive to HTS surface degradation. The polishing of the surface leads to appearance of a dynamic effect which is peculiar to compact vortex structures considered above and consists in a dynamic surmounting of the surface barrier by penetrating vortices.

## 5. Applications

Using the obtained results we have utilized the levitation systems in many practical methods and devices: the methods for energy loss measurements in II-type superconductors [77, 78] and any electromagnetic materials [79], the method for measuring magnetic field gradients [80], viscosimeter [81, 82], accelerometer [83], vibrometer [84], the method of superconducting volume fraction determination [85], the methods for critical current density evaluation [19, 86], the method for for quality estimation of superconducting joints in bulk melt-processed high temperature superconductors [21, 22], the high-speed magnetic rotor [52, 53], the contactless bearing with self-centering and intrinsic damping of parasitic vibrations [51].

The last one uses both melt-textured and granular HTS components and accumulate the knowledge of the results which have been reviewed here. Scheme of such a bearing is shown Fig. 18 (left). The melt-processed components have been used as a main carrying system to support the weight of the rotor and to provide the low AC losses at its rotations. The granular HTS were used to obtain the self-centering effect during cooling and to decrease the parasitic resonances in the system [13]. The AC loss dependencies on the rotation speed of the rotor in the systems with and without the granular component are also shown in Fig. 18 (right). The granular HTS decreases the resonance peak and increases the rotor stability. Such effective damping of the low frequency vibrations is a consequence of high energy losses due to intragrain magnetic flux motion in these materials, but only for well optimized rotors, when AC field is less than $b_c$ the contribution of these granular components to energy loss at high frequency is negligible [32, 51].

## Conclusions

In this paper we have presented a short review of the experimental techniques with levitation such as levitation force measurements, the resonance oscillation technique, the methods based on contactless magnetic rotors, etc. These techniques are non-destructive by nature, easy to use and allow to obtain unique information about the physical properties of the HTS with high accuracy. We have used these techniques to investigate the macroscopic magnetic properties of the granular and quasi-single crystal melt-processed HTS, and fundamental properties of compact vortex structures dynamics in HTS bulks.

It has been proved that the surface barrier plays an important role forming the macroscopic magnetic properties of HTS ceramics. At small AC field amplitudes (< 15 G), it prevents the magnetic flux penetration into superconducting grains. At bigger amplitudes it causes the hysteretic energy loss. It is the surface barrier rather than bulk pinning that is responsible for the range of quasi-equilibrium positions in the PM–HTS systems and for the short time magnetic relaxation in Bi-ceramics. In the field cooled regime, the surface barrier is the main reason for the flux repulsion and the PM floating-up effect from above the HTS ceramics.

The intragrain vortex dynamics is determined by weak ($U_p \sim kT$) pinning centers, like oxygen vacancies. This demonstrates that the strong ($U_p \gg kT$) pinning centers, like dislocations and twin boundaries, are practically absent in the superconducting grains of the HTS ceramics made by the standard sintering techniques. Three regions of vortex motion in the Lorenz force scale (AC field amplitude) have been determined and explored: thermally assisted flux motion (TAFF), thermally activated flux creep (TAFC) and free flux flow (FF). TAFF and FF are linear, i.e. can be assigned by constant viscosity of vortex motion.

The vortex dynamics in melt-processed quasi-single crystals of Y-123 has essentially hysteretic nature and can be well described by the critical state model with a strong pinning mechanism. The whole physics of the PM–HTS interaction is concentrated in a thin undersurface layer of HTS samples where the compact vortex structures alternately penetrates into HTS during PM oscillations. The surface barrier, which can be observed at specially polished surfaces, produces a dynamic effect the nature of which is peculiar to compact vortex structures behavior under influence of a strong pinning. The dislocations are considered to be the most probable candidates for such a strong pinning.